# Effect of solute content and temperature on the deformation mechanisms and critical resolved shear stress in Mg-Al and Mg-Zn alloys

J-Y. Wang[a, b], N. Li[a, b], R. Alizadeh[a], M. A. Monclús[a], Y. W. Cui[c], J. M. Molina-Aldareguía[a], J. LLorca[a, b,*]

[a]IMDEA Materials Institute, C/Eric Kandel 2, 28906, Getafe, Madrid, Spain
[b]Department of Materials Science, Polytechnic University of Madrid/Universidad Politécnica de Madrid, E.T.S. de Ingenieros de Caminos, 28040 Madrid, Spain
[c]School of Materials Science and Engineering, Nanjing Tech University, Nanjing 210009, P. R. China.

**Abstract**

The influence of solute atoms (Al and Zn) on the deformation mechanisms and the critical resolved shear stress for basal slip in Mg alloys at 298 K and 373 K was ascertained by micropillar compression tests in combination with high-throughput processing techniques based on the diffusion couples. It was found that the presence of solute atoms enhances the size effect at 298 K as well as the localization of deformation in slip bands, which is associated with large strain bursts in the resolved shear stress ($\tau_{RSS}$)-strain (ε) curves. Deformation in pure Mg and Mg alloys was more homogeneous at 373 K and the influence of the micropillar size on the critical resolved shear stress was much smaller. In this latter case, it was possible to determine the effect of solute content on the critical resolved shear stress for basal slip in Mg-Al and Mg-Zn alloys.

# 1. Introduction

Mg and its alloys stand for the lightest structural metals and present high specific-strength, excellent bio-compatibility and reduced cost [1–5]. One main limitation for the engineering application is the reduced ductility and formability at room temperature [1] whose origin can be traced to their low-symmetry hexagonal closed packed (hcp) lattice structure. Dislocation slip in hcp Mg crystals mainly occurs by basal and prismatic slip along <a> directions as well as by <c+a> dislocations on the pyramidal planes. The critical resolved shear stress (CRSS) for dislocation slip in the basal $(0001)<11\bar{2}0>$ system is very low (~0.5 MPa for pure Mg [6–8]) as compared with the CRSS values for pyramidal and prismatic slip (~ 40 MPa for pure Mg [9–11]). According to the von Mises criterion [12], the general homogeneous plastic deformation in polycrystals can be only accommodated by five independent slip systems. As basal slip can only provide two independent slip systems [13,14] and the CRSS value for pyramidal slip is very large, deformation along the <c> axis has to be accommodated by twinning, a polar mechanism which only takes place when the $c$ axis of the hcp lattice is extended. Thus, the large differences in the CRSS values to activate plastic slip in the different systems in Mg as well as the polarity of twinning lead to the large plastic anisotropy of Mg alloys, that has very negative effects on the ductility.

The ductility of Mg alloys can be improved by reducing the differences in the CRSS values between the basal and non-basal slip systems through the strengthening of basal slip or softening of pyramidal slip via solid solution. Moreover, the overall yield strength of Mg alloys is improved by increasing the CRSS for basal slip. Solid solution strengthening is a convenient approach to achieve these goals and the influence of different solute atoms (Zn, Cd, Sn, In, Al, Pb, Bi, Y and Dy) on the CRSS values for basal slip has been studied experimentally [15–21]. For instance, tensile tests in single crystal Mg-Al [16] and Mg-Zn [17] alloys showed an increase of the CRSS values for basal slip up to 2.5 MPa and 2 MPa with the addition of 1.6 at.%Al and 0.5 at.%Zn, respectively. The solid solution strengthening of basal slip due to Y and Dy was measured through compression tests in single crystals, and the CRSS increased up ~10 MPa and ~5 MPa with the addition of 1.0 at.%Y and 0.5 at.%Dy, respectively [15]. Nevertheless, reliable experimental data of the influence of solute hardening on the CRSS values for basal slip are scarce because the mechanical testing of single crystals is very time consuming, as single crystals with various solute contents have to be manufactured. Moreover, it is difficult to obtain the CRSS values for basal slip from mechanical tests in polycrystalline samples because of the superposition of different strengthening contributions (grain boundaries, latent hardening, texture) that cannot be easily isolated [22].

Micromechanical testing techniques, such as nanoindentation and micropillar compression, offer an attractive approach to obtain reliable values of the CRSS for basal slip as a function of the solute content. The main problem of these techniques is associated with the overestimation of the strength when the volume of the tested material is of the order of tenths of μm$^3$ or smaller [23–27], due to the limited number of mobile dislocations and the scarcity of dislocation sources in the deformed volume. For instance, the CRSS for basal slip measured by K. Eswar Prasad et al. [25] in pure Mg micropillars of 3 μm in diameter was six times higher than the one measured in square specimens of 3 x 3 mm$^2$ cross-section. Another study focused in the compression of micropillars of pure Mg with different diameters concluded that micropillars of at least 10 μm in diameter should be used to avoid size effects [24]. Due to these limitations, nobody has tried to determine the influence of the solute content on the CRSS for basal slip in Mg in the past. More recently, Nayyeri et al. [21] tried to solve this problem by extrapolating the CRSS using spherical nanoindenters of different radius to an infinite radius indenter. The results for Mg-Al and Mg-Zn alloys were in agreement with previous data in the literature for single crystals [16,17].

In this investigation, micropillar compression tests in combination with high-throughput processing techniques based on the diffusion couples [28–31] were used to ascertain the influence of solute atoms on the CRSS for basal slip in Mg-Al and Mg-Zn alloys at 298 K and 373 K. Micropillars with different dimensions were manufactured by the focus ion beam (FIB) technique and the size effects (and the associated deformation mechanisms) were analyzed in terms of solute content and temperature. It was found that the size effects impede the determination of the actual values of the CRSS at 298 K, but not at 373 K, because of the higher dislocation mobility that reduces the influence of the micropillar dimensions on the deformation mechanisms.

## 2. Materials and experimental techniques

### 2.1 Materials

Pure Mg, Mg-9 at.%Al, Mg-1.5 at%Zn and Mg-2.5 at%Zn alloys were manufactured by casting in an induction furnace (VSG 002 DS, PVA TePla) under a protective Ar atmosphere to avoid oxidation from high-purity Mg (99.99 wt.%), Al (99.99 wt.%) and Zn (99.99 wt.%) pellets. The cast ingots were homogenized in an Ar atmosphere into quartz capsules at 673 K for 15 days to achieve a homogeneous composition and induce grain growth. Afterwards, they were cut into discs of 12 mm in diameter and 7.5 mm in length. The surfaces of the pure Mg, Mg-9 at.%Al and Mg-2.5 at.%Zn alloy discs were mirror-polished using standard metallographic techniques and diffusion-bonded in vacuum under a compression force of 800 N for 1 h at 673 K in the thermomechanical simulator Gleeble 3800, to produce Mg/Mg-9 at.%Al and Mg/Mg-2.5 at.%Zn diffusion couples (referred to as Mg-Al and Mg-Zn diffusion couples, respectively, in the rest of the paper for the sake of clarity). Subsequently, the diffusion couples were further annealed into quartz capsules in an Ar atmosphere at 673 K during 352 h and 336 h, for the Mg-Al and the Mg-Zn diffusion couples respectively, to promote the diffusion of the solute atoms across the interface.

*2.2 Microstructure characterization*

Parallelepipedal specimens of 7 x 7 x 2 mm$^3$ were cut at low speed to minimize surface deformation from the Mg-Al and Mg-Zn diffusion couples (with the shortest dimension parallel to the interface) and from the Mg-1.5 at.%Zn billet. The 7 x 7 mm$^2$ surface was polished using abrasive SiC papers with a grit size of 320, 600, 1200 and 2000. Afterwards, the samples were polished using a MD-Mol cloth with a 3 μm diamond paste, followed by polishing with the MD-Nap cloth with 0.25 μm diamond paste. Finally, the samples were chemically polished using a solution of 75 ml ethylene glycol, 24 ml of distilled water and 1 ml of nitric acid to remove any residual surface damage and to reveal the grain boundaries.

The composition profile along lines perpendicular to the interface were measured in the diffusion couples by electron probe microanalysis (EPMA) using Wavelength Dispersive Spectroscopy (WDS) with a voltage of 20 kV and a beam current of 50 nA in a JEOL Superprobe JXA-8900M. In addition, the microstructure of the samples was analyzed in a dual beam scanning electron microscope (Helios Nanolab 600i FEI) using electron backscatter diffraction (EBSD) to identify the grain orientation and the grain size. The EBSD measurements were carried out at an accelerating voltage of 30 kV and a beam current of 2.7 nA using a step size of 1.5 μm.

*2.3. Micromechanical characterization*

Fig. 1(a) and Fig. 2 show the representative EBSD maps at the interface of Mg-Al and Mg-Zn diffusion couples, respectively. The grain sizes were very large (around 1 mm) and single-crystal micropillars for the mechanical characterization could be easily machined using FIB in the center of grains, guided by the EBSD information, far from any grain or twin boundaries. In the case of the diffusion couples, the Al and Zn content changed smoothly within the diffusion region, as shown in Fig. 1(b) and Fig. 2 where the gray dots represent the composition profile of each alloying elements, Al and Zn as determined by EPMA. The micropillars were milled in grains suitably oriented for basal slip with the following compositions: pure Mg, Mg-4 at.%Al, Mg-9 at.%Al, Mg-1.5 at.%Zn and Mg-2 at.%Zn. In the case of the diffusion couples, all the micropillars for a given composition were milled in a very small region (smaller than 15 x 15 $\mu$m) to ensure that the composition was constant.

The Schmid factors (SF) corresponding to basal slip and tensile twinning under compression were computed [32] in each case, based on the crystal orientations obtained from the EBSD measurements. They are depicted in Table 1 for the selected grains with different chemical composition. The high SF values for basal slip (> 0.45) ensured that the micropillars were favourably oriented for basal slip.

Micropillars with a square cross-section were machined to facilitate the observation and analysis of the slip traces on the lateral surfaces after deformation (Fig. 3). The milling process was carried out using a $Ga^+$ ion beam with an accelerating voltage of 30 KV. A beam current of 9.3 nA was used for the coarse milling step and an ion beam current of 40 pA was employed in the final polishing step to minimise the surface damage due to $Ga^+$ ion-implantation. Micropillars with cross-sections of 3 x 3 $\mu m^2$, 5 x 5 $\mu m^2$ and 7 x 7 $\mu m^2$ were machined inside each grain of interest. The final taper of the micropillars was below 1º and the aspect ratio was in the range 2:1 to 3:1 in all cases to avoid buckling for higher aspect ratios, or non-uniform stresses along the length for lower aspect ratios [33,34]. The micropillars were observed both before and after the deformation using high resolution scanning electron microscopy (HR-SEM) to ensure that they were free of defects and to determine the deformation micro-mechanisms.

Micropillar compression tests were firstly carried out within the scanning electron microscope (FEI Helios Nanolab 600) at 298 K (25 ℃) using a Hysitron PI87 SEM Picoindenter. The *in-situ* tests were used to observe the formation of the slip traces along the surface of the pillars. The misalignment between the tip and the top surface of the pillars was carefully corrected in these tests. Micropillar compression tests were also performed at 373 K (100 °C) using a TI950 Triboindenter (Hysitron, INC., Minneapolis,

MN). For the 100 °C tests, a hot stage (Hysitron xSol) and a flat punch fitted to a special long insulating shaft were used. In this configuration, the sample is placed between two resistive heating elements in order to eliminate temperature gradients across the sample thickness. Dry air and argon were used to purge the testing area around the tip and sample surface to prevent oxidation. Once the sample reached 100 ºC, the tip was placed at about 100 μm from the pillar surface for 10-15 minutes, to ensure passive heating of the tip before the start of the test and minimize thermal drift.

Load was applied in all cases through a diamond flat punch with 10 μm in diameter. All the tests were implemented under displacement-control mode (which was achieved through a feedback control system) at an average strain rate of $10^{-3}$ s$^{-1}$ up to a maximum strain of 10%. At least three tests were repeated for each condition (composition, size and temperature) to ensure the reproducibility of the results. The experimental load-displacement curves were primarily corrected to account for the compliance due to the elastic deflection of the base of the pillar, following the Sneddon correction [35,36]. Afterwards, the corrected curves were converted to engineering stress (σ)-strain (ε) using the initial cross-sectional area and the height of the micropillars measured after the milling process. As the micropillar compression tests were performed in grains with different orientations, the results of the mechanical tests were plotted using the engineering resolved shear stress for basal slip, $\tau_{RSS} = SF \times \sigma$, where *SF* stands for the Schmid factor for basal slip in each case (Table 1).

The deformed micropillars were carefully examined using secondary electrons in the HR-SEM to determine the orientation of the slip traces that appeared on the free surfaces. In addition, thin lamellas of ≈ 100 nm in thickness were trenched and lifted-out from selected deformed micropillars by FIB. They were subjected to transmission Kikuchi diffraction (TKD) to elucidate the crystal orientation after deformation and to ensure that the deformed micropillars did not suffer any twinning. The electron beam was operated at 15 kV with a beam current of 2.7 nA and the TKD maps were collected with a step size of 70 nm.

## 3. Experimental results

### *3.1 Mechanical response at room temperature (298 K)*

Representative resolved shear stress ($\tau_{RSS}$)-strain (ε) curves corresponding to micropillars with different sizes tested at room temperature for Mg, Mg-Al and Mg-Zn alloys are plotted in Fig. 4 and Fig. 5. The results of the mechanical tests were consistent and the scatter was limited. The slope corresponding to the initial elastic region was always slightly lower than that associated with elastic unloading at the end of the test, regardless of the micropillar diameter. These differences could be attributed to the impossibility of attaining a completely perfect parallelism between the top surface of the pillar and the flat punch [37], even though large efforts were made to correct for the misalignment during the *in situ* tests. Despite this, the initial plastic yielding could be easily identified as discontinuities or changes in the initial slope in the stress-strain curves in all cases.

Regardless of the solute content, a pronounced size effect of the type "smaller is stronger" was found in all cases, in agreement with previous investigations [23,24]. The flow stress of the 3x3 μm$^2$ micropillars was always substantially higher than that of larger micropillars, but the discrepancies in the flow stress between 5x5 μm$^2$ and 7x7 μm$^2$ were much smaller. However, substantial differences were found in the shape of the $\tau_{RSS}$-ε curves as a function of solute content and pillar size, which indicates differences in the way the pillars deformed, as described in detail below.

In the case of pure Mg (Fig. 4a), basal slip was activated at very low stresses as stated above. This is not surprising considering the low CRSS for basal slip in pure Mg. Nevertheless, due to the annealing condition, the initial dislocation density was very low and it is foreseen that the density of mobile dislocations will exhaust soon with the applied strain as dislocations scape out at the micropillar surface. As a result, the initial yield is followed by a hardening stage for ε < 1% that is more pronounced the smaller the pillar size, especially for the 3x3 μm$^2$ micropillar. When the applied strain reaches 2%, strain bursts begin to appear in the resolved shear stress ($\tau_{RSS}$)-strain (ε) curves giving rise to the jerky behaviour typically found in micropillars. They are represented by sudden drops in stress due to the loss of contact between the flat punch and the micropillar, and result from the sudden activation of dislocation sources at higher stresses. The magnitude of the strain bursts increases with the magnitude of the resolved shear stress (and thus, with the reduction in micropillar size), leading to large strain jumps in the micropillars with lateral dimensions of 3 and 5 μm, while the 7 μm micropillar displays a much smoother behaviour. These observations agree well with the accepted theory of dislocation source exhaustion [38].

The orientation of the slip traces indicated the activation of basal slip in all cases (and the absence of twinning was confirmed by transmission Kikuchi diffraction, see supplementary material), but the morphology of the slip traces differed with pillar size. Most of the plastic deformation in the 3 x 3 μm² micropillar of Fig. 6a is concentrated in one slip band which corresponds to the large strain jump in Fig. 4a. Deformation in the 5 x 5 μm² micropillar (Fig. 6b) is localized in two slips bands (one more predominant than the other). However, the surface of the largest 7 x 7 μm² micropillar presents many slips traces (Fig. 6c), homogeneously distributed along the length of the pillar, in agreement with a more continuous, less jerky behaviour of the $\tau_{RSS}$-ε curves, as depicted in Fig. 4a. Nevertheless, the magnitude of the size effect between the 5 x 5 μm² and 7 x 7 μm² micropillars was negligible. These observations indicate that the behaviour of the 7 x 7 μm² micropillar is less affected by the scarcity of dislocation sources and hence should be close to be free of pillar size effects and resemble the bulk behaviour.

Similar size effects were also found in the Mg-Al (Fig. 4) and Mg-Zn (Fig. 5) alloys, but there were significant differences in the shape of the resolved shear stress ($\tau_{RSS}$)-strain (ε) curves. For instance, contrary to pure Mg, a clear initial elastic region was found in the ($\tau_{RSS}$)- (ε) curves for Mg-Al and Mg-Zn alloys. This behaviour indicates that the few dislocations that are initially present in the micropillars are strongly pinned by the solute atoms and were not able to contribute to the plastic slip, especially for the higher solute contents. So, the behaviour of Mg-4 at%.Al (Fig. 4b) was similar to that of pure Mg, but for higher Al contents (Fig. 4c) and Mg-Zn (Fig. 5), the elastic region finished abruptly with the development of large strain bursts. The strain bursts were bigger the smaller the pillar size, indicating that, contrary to pure Mg, initial yielding in the Mg alloys is accomplished by the activation of single-arm dislocation sources. This behaviour is in good agreement with the SEM micrographs of the deformed micropillars in all cases, for instance, those shown in Fig. 7 for Mg-9 at.%Al, that indicate that plastic deformation localizes in a few slip bands for all micropillar diameters. The movies of the *in-situ* micropillar compression tests for the 3 μm and 7 μm micropillars in Mg-9 at.%Al (see the supplementary material section) clearly show that the strain jumps at strains between 1% and 7% are associated with the sudden formation of slip bands (shown in Fig. 7a) that appeared simultaneously in case of the small 3x3 μm² micropillar. Plastic deformation in the 7x7 μm² micropillar was also discontinuous and took place with the progressive formation of slip bands of different intensity corresponding to the successive strain bursts, as can be seen in the deformed pillar of Fig. 7c. The behaviour of the 5x5 μm² micropillar was intermediate between those found for the 3x3 μm² and 7x7 μm² micropillars.

All in all, the described behaviour agrees well with the expected behaviour for Mg micropillars favourably oriented for basal slip, but hinders the determination of solid solution strengthening effects from the tests. While it is possible to determine a yield stress for the Mg (Fig. 4a) and Mg- 4 at.%Al (Fig. 4b) micropillars before the jerky behaviour (characteristic of the operation of single arm dislocation sources) sets on, this is impossible for Mg-9 at.% Al (Fig. 4c) and Mg-Zn (Fig. 5), for which plasticity directly triggers strain bursts associated with the activation of dislocation sources. It is speculated that the few mobile dislocations that are initially present in the micropillars are strongly pinned by the solute atoms and are not able to glide at room temperature before the jerky behaviour sets on. This behaviour could potentially be reverted by thermal activation, as shown below.

### 3.2 Mechanical response at 373K (100 °C)

The resolved shear stress ($\tau_{RSS}$)-strain (ε) curves corresponding to micropillars with different cross sections (5 x 5 μm² and 7 x 7 μm²) tested at 373 K (100 °C) for Mg-Al and Mg-Zn alloys are presented in Fig. 8 and Fig. 9, respectively. The scatter was limited. The general features of the resolved shear stress ($\tau_{RSS}$)-strain (ε) curves at 373 K (100 °C) were similar to those observed at ambient temperature, but some differences could be found in the way plasticity developed, especially for the Mg alloys with the higher solute content. In the case of pure Mg (Fig. 8a) and Mg-4 at.%Al (Fig. 8b), temperature has a negligible effect and the initial yield stress, the flow stresses at large plastic strains and the magnitude of the strain bursts for micropillars of 5x5 μm² and 7x7 μm² are identical to those found at ambient temperature (Fig. 4a and 4b). The deformed micropillars also showed a strong localization of the deformation in slip bands at 373 K (100 °C), as shown in Fig. 10.

The $\tau_{RSS}$- ε curves at 373 K are plotted in Fig. 8c in the case of Mg-9 at.%Al, Fig. 9a for Mg-1.5 at.%Zn and Fig. 9b for Mg-2 at.%Zn alloys. The curves at 373 K (100 °C) also tend to reach a plateau in flow stress at large applied strains, at levels similar to those found at room temperature, but with a negligible size effect if the 5x5 μm² and 7x7 μm² micropillars are compared. Moreover, the elastic to plastic transition is remarkably different to that found at room temperature (Fig. 4c and Fig. 5). First, the magnitude of the strain bursts is much smaller than that found at ambient temperature for the same micropillar size. Second, and as a result of the first, the initial yielding is much more gradual than that found at room temperature. Third, the deformation along the micropillars should be more homogeneous than that found at room temperature, and this hypothesis is supported by the SEM micrographs of the 5x5 μm$^2$ and 7x7 μm$^2$ Mg-9 at.%Al micropillars shown in Fig. 11a and 11b, respectively. Finally, the $\tau_{RSS}$-ε curves presented in Fig. 8 show evidence of serrated flow that can be attributed to the dynamic interaction between solute atoms and dislocations. This mechanism (also named dynamic strain ageing) has also been reported for binary Mg alloys tested at high temperature [39,40], but not at room temperature [41], except for nanostructured binary Mg-Al alloys [42]. The results indicate that thermal activation helps overcome the pinning of the dislocations by the solute atoms in Mg-9 at.%Al and Mg-Zn, so that after the initial elastic loading, yielding takes place smoothly by the gliding of pre-existing basal dislocations, followed by the onset of the typical jerky behaviour and the activation of single arm dislocation sources at larger stresses, similarly to what happens in pure Mg and Mg-4 at.%Al at room temperature. To further ascertain the dominant deformation mechanisms at 373 K, a thin foil from the deformed 7x7 μm$^2$ micropillar tested at 373K (100ºC) in pure Mg was analysed by the transmission EBSD, as illustrated in the Supplementary Fig. 1. Basal slip was the only active mechanism without any contribution from twinning. Moreover, the analysis of a thin foil extracted from a Mg-9 at.%Al micropillar tested at 373 K by TEM verified that the micropillars were free of precipitates after high temperature testing. It should be finally noted that the analysis of the lateral and top surfaces of the micropillars in the SEM before and after testing did not show any traces of oxidation.

## 4. Discussion

### *4.1 Factors influencing the initial plastic yielding of Mg alloy micropillars favorably oriented for basal slip*

The results presented above indicate that the initial yielding of Mg alloy micropillars favorably oriented for basal slip depends on a competition between the mobility of the pre-existing dislocations and the stress required to activate single-arm dislocation sources, which in turn depends on micropillar size and testing temperature. For low alloying contents (pure Mg and Mg-4 at.%Al), initial plastic yielding is smooth and progresses through the gliding of pre-existing dislocations at very low CRSSs at 273 K and 373 K, As the pre-existing dislocations scape out of the pillar surface for the smaller pillars (3x3 μm$^2$), further elastic loading is required to the point at which large strain bursts start to develop associated with the activation of single arm dislocation sources, in agreement with the results in the literature [23,24]. Micropillars larger than 5x5 μm$^2$ are, however, less affected by size effects and display a smoother stress-strain curve. The observed behavior was different for high Al contents and for Mg-Zn. In this case, the pre-existing dislocations are strongly pinned by the solute atoms and initial plastic yielding at room temperature is dominated by the activation of single arm dislocation sources even for the largest 7x7 μm$^2$ micropillar. Increasing the testing temperature up to 373 K, however, helps overcome the pinning of the dislocations by the solute atoms in Mg-9 at.%Al and Mg-Zn, so yielding takes place smoothly by the gliding of pre-existing basal dislocations after the initial elastic loading, reducing the magnitude of the size effect in 5x5 and 7x7 μm$^2$ micropillar.

If this behavior is a result of the low initial density of mobile dislocations within the well-annealed pillars, a higher initial dislocation density should induce a less jerky behavior for the same pillar size at room temperature. This hypothesis was tested in a 5x5 μm$^2$ Mg-9 at.%Al micropillar by pre-straining it at 373 K up to a strain of 2% to increase the dislocation density and re-testing it again at ambient temperature. The corresponding $\tau_{RSS}$-ε curves are plotted in Fig. 12 and compared with that obtained at room temperature for Mg-9 at.%Al without pre-straining. Indeed increasing the initial dislocation density reduced the magnitude of the strain bursts during deformation at room temperature and the resolved shear stress to deform the micropillar, confirming the initial hypothesis. However, deformation was still jerky and localized in a few shear bands (see inset in Fig. 12). Thus, the initial dislocation density in the pre-strained pillar was not large enough to completely erase size effects at room temperature in Mg- 9 at.%Al.

*4.2 Determination of the effect of solute strengthening in the CRSS for basal slip*

Representative $\tau_{RSS}$-ε curves at the ambient temperature and 373 K (100 ºC) of the micropillars with 7 x 7 μm$^2$ cross section are plotted in Figs. 13a and 13b, respectively, for all the alloys to show more clearly the effect of the solute atoms on the mechanical response. The micropillar compression tests carried out at ambient temperature (Fig. 13a) show the strengthening effect of the solute atoms, but it is not possible to determine the effect of the solute content on the actual CRSS because of the size effects encountered as explained in the previous section.

The size effect on the micropillar tested at 373 K (Fig. 13b) was significantly smaller than that found at ambient temperature, regardless of the solute content. In fact, plastic deformation was continuous from the beginning of the deformation for Mg and Mg alloys in the micropillar with 7 x 7 μm² cross section, and a clear yield point could be identified at the beginning of the $\tau_{RSS}$-ε curves in all cases. Even though the CRSS determined for pure Mg at the yield point (≈ 7 MPa) for micropillars with lateral dimensions of 5 and 7 μm was higher than the value reported in the literature (below 1 MPa [6–8]), it seems reasonable to determine the influence of the solute content on the CRSS at 373 K as the difference in the CRSS at the initial yield point between the studied alloys and the pure Mg, Δτ. The corresponding results are plotted in Fig. 14 for Mg-Al alloys and Mg-Zn alloys for the micropillars with lateral dimensions of 7 μm.

In order to validate the solute strengthening effects determined using micropillar compression, it is instructive to compare the results with previous works that used conventional bulk testing. The investigations in the past to determine the CRSS for basal slip in Mg-Al and Mg-Zn were carried out by means of tensile tests of single crystals [16,17]. The results are plotted in Fig. 14 for the sake of comparison. Both data at 25 ºC and 100 ºC are plotted in each case but they are indistinguishable because solid solution strengthening in Mg enters the athermal regime above room temperature, [16,17]. Unfortunately, past fundamental studies in single crystals only covered dilute alloys, i.e., concentrations of less than 0.5 at.% in alloying content, within the range of solubility at room temperature. An exception is the work of Chun and Byrne [43], which determined the CRSS for basal slip in supersaturated Mg-2 at.%Zn single crystals, and the results are also plotted in Fig. 14b. Results for single crystals were not found for supersaturated Mg-Al alloys, but Cáceres and Rovera [41] estimated the CRSS for basal slip from polycrystalline Mg-Al alloys and their data is also plotted in Fig. 14a for the sake of comparison.

The strengthening due to the solid solution can be modeled using the Labusch model [44–46] according to:

$$\Delta\tau = \tau_{alloy} - \tau_{Mg} = Kc^{2/3} \qquad (1)$$

where c is the solute element content (in at. %) and K is a constant that depends on the alloying element and temperature [44,46]. The experimentally determined strengthening in CRSS at 373 K of Fig. 14a and 14b correspond to K = 1.4 MPa$^{-1}$ and K = 3.48 MPa$^{-1}$, for supersaturated Mg-Al and Mg-Zn, respectively. In the case of Mg-Al alloys, the results are in good agreement with the work of Cáceres and Rovera [41] and interestingly the strengthening slope K is the same as that for diluted Mg-Al alloys [16]. This suggests that random solid solution effects, which control the strengthening of dilute Mg–Al alloys, can be extrapolated to the supersaturation regime. In the case of Mg-Zn alloys, however, the current results are in good agreement with previous bulk data [43], but the strengthening slope K in the supersaturated regime is much larger than that found for diluted alloys [17]. Therefore, the evidence suggests that even though Mg and Al have similar strengthening effects for dilute alloys, Zn is far more efficient than Al in strengthening Mg at high solute concentrations. Among others, the tendency of Zn atoms for solute clustering [43] and short range order [47] have been considered as possible reasons for this behavior.

## 5. Conclusions

An experimental methodology based on the diffusion-couples and compression of micropillars has been developed to assess the influence of solute content and temperature on the CRSS values for the basal slip in Mg alloys. The Mg-MgAl and Mg-MgZn diffusion couples (as well as the Mg-Zn alloy) were manufactured by casting followed by long term annealing at high temperature. Micropillars of lateral dimensions in the range 3-7 μm were carved by focused ion milling in grains favorably oriented for basal slip and deformed in compression at 298 K and 373 K.

The micropillar compression tests at 298 K showed a large size effect (the smaller the stronger) associated with the limited number of mobile dislocations and the activation of single arm dislocation sources. The size effect led to the development of large strain bursts in which deformation was localized in very few slips bands. These mechanisms were more noticeable as solute content increased, because of the pinning effect of solute atoms on dislocations, and, as a result, it was not possible to obtain size-independent values of the influence of the solute content on the critical resolved shear stress for basal slip.

Size effects were much smaller in micropillars tested at 373 K. Moreover, strain bursts almost disappeared for micropillars with a cross-section of 7 x 7 μm$^2$ and deformation by basal slip was fairly homogeneous along the micropillar length. These changes were attributed to the higher dislocation mobility at 373K. Under these circumstances, it was possible to estimate the contribution of the solute atoms to the critical resolved shear stress as the difference in the shear stress at the initial yield point between the alloys and the pure Mg micropillars. It was found that the solid solution strengthening provided by Al and Zn at 373 K followed the Labusch model and was in good agreement with previous experimental data obtained by bulk compression of single crystals.

Overall, the experimental observations presented in this paper highlight the experimental difficulties to extract quantitative, size-independent values of the critical resolver shear stress for pure metals and alloys using micromechanical testing techniques (such as micropillar compression or nanoindentation). It confirms (in agreement with previous observations [48-50]) that the difficulties associated with the size effect decrease as the testing temperature increases due to higher mobility of the dislocations and that these techniques may be particularly suitable to determine the mechanical properties of single crystals at high temperature.

**Acknowledgements**

This investigation was supported by the European Research Council (ERC) under the European Union's Horizon 2020 research and innovation programme (Advanced Grant VIRMETAL, grant Agreement no. 669141). Ms. J-Y. Wang and Ms. N. Li acknowledge the financial support from the China Scholarship Council (Grants no. 201506890002 and 201506020081, respectively). R. Alizadeh also acknowledges the support from the Spanish Ministry of Science through the Juan de la Cierva program (FJCI-2016-29660).

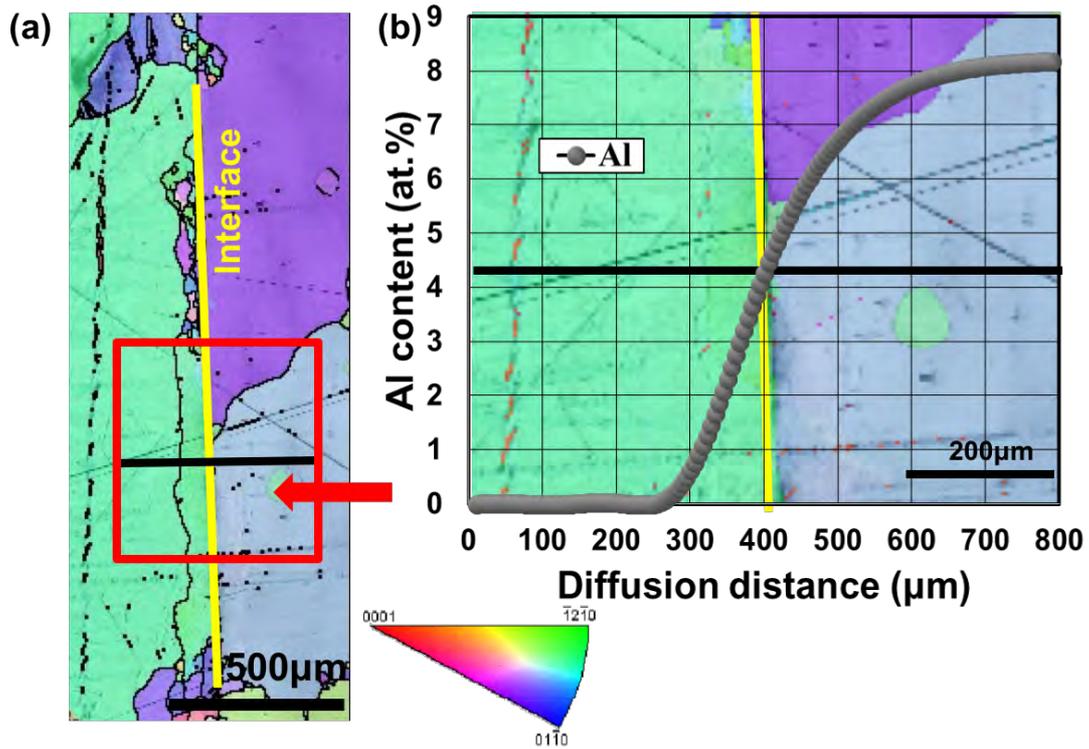

Fig.1 (a) The orientation measurement of the Mg-MgAl diffusion region in by EBSD; (b) Composition profile (measured by EPMA) as function of the diffusion distance together with the crystal orientation (measured by EBSD) given by the inverse pole figure (IPF) in the Mg-Al diffusion couple. The yellow line indicates the interface and the black line stands for the position where the composition profile was measured.

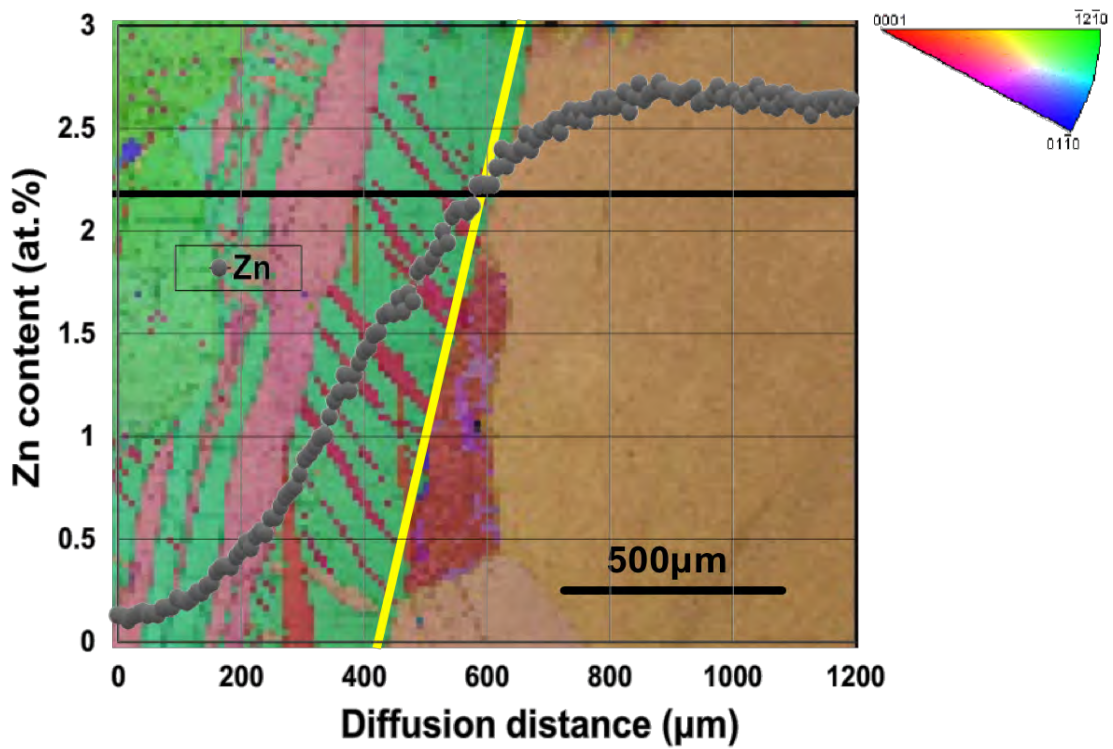

Fig.2 Composition profile (measured by EPMA) as function of the diffusion distance together with the crystal orientation (measured by EBSD) given by the inverse pole figure (IPF) in the Mg-Zn diffusion couple. The yellow line indicates the interface and the black line stands for the position where the composition profile was measured.

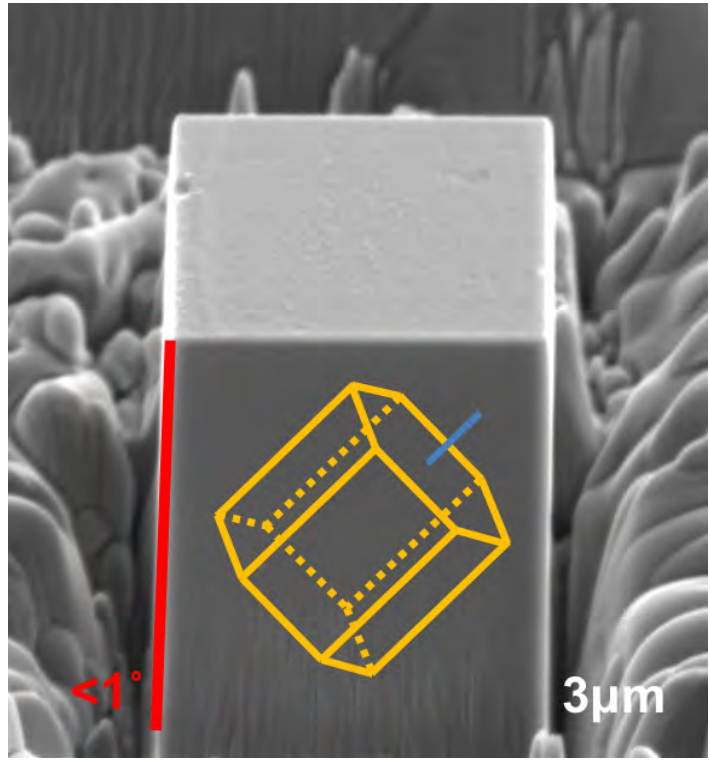

Fig. 3. Scanning electron micrograph of the micropillar with a square section of 7 x 7 µm$^2$ and taper < 1°. The insert shows schematically the orientation of the HCP lattice to show that the micropillar is favourably oriented for basal slip.

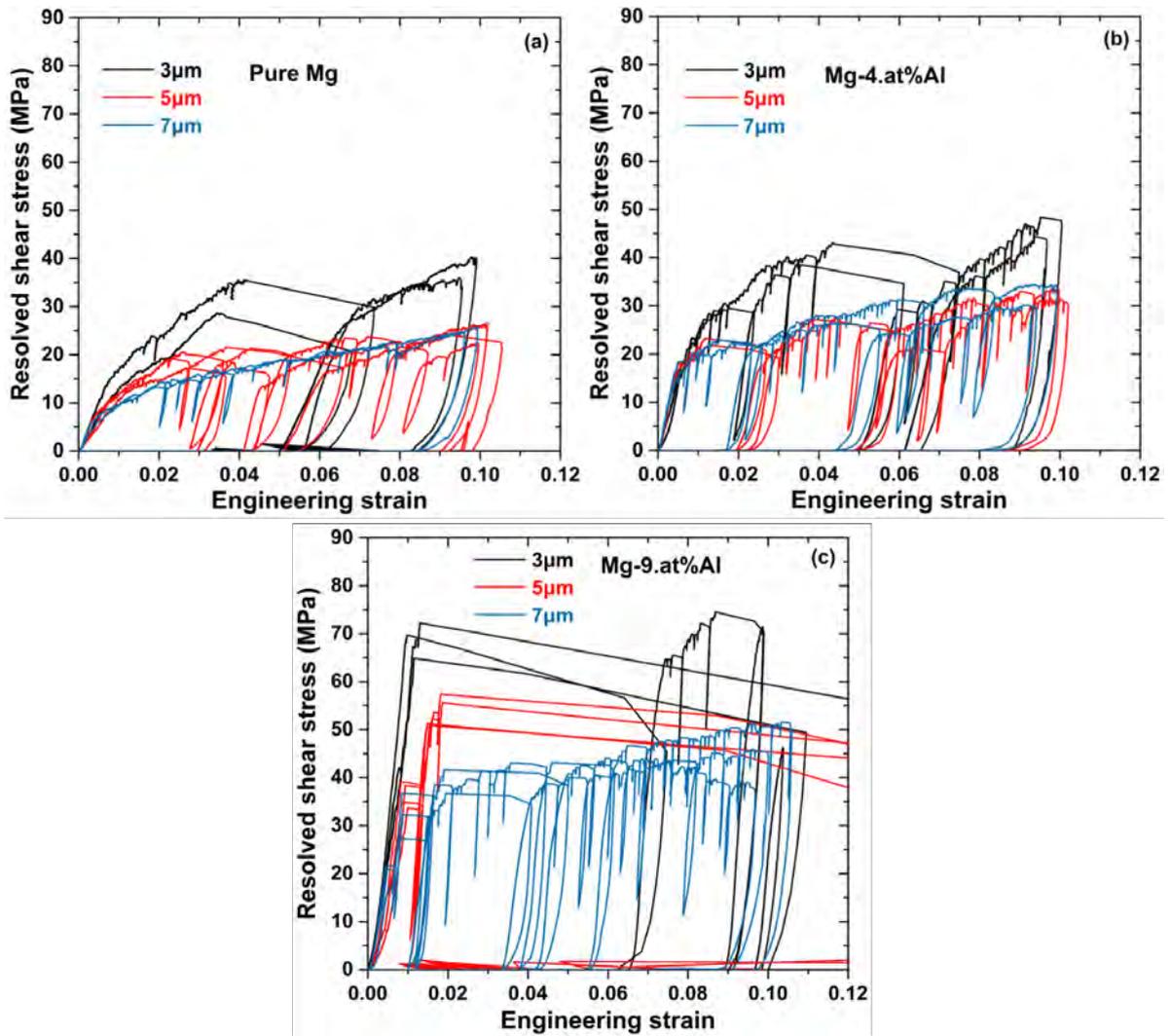

Fig. 4. Resolved shear stress ($\tau_{RSS}$) vs. engineering strain (ε) curves for micropillars of different dimensions tested at ambient temperature. (a) Pure Mg. (b) Mg-4 at.%Al. and (c) Mg-9 at.%Al.

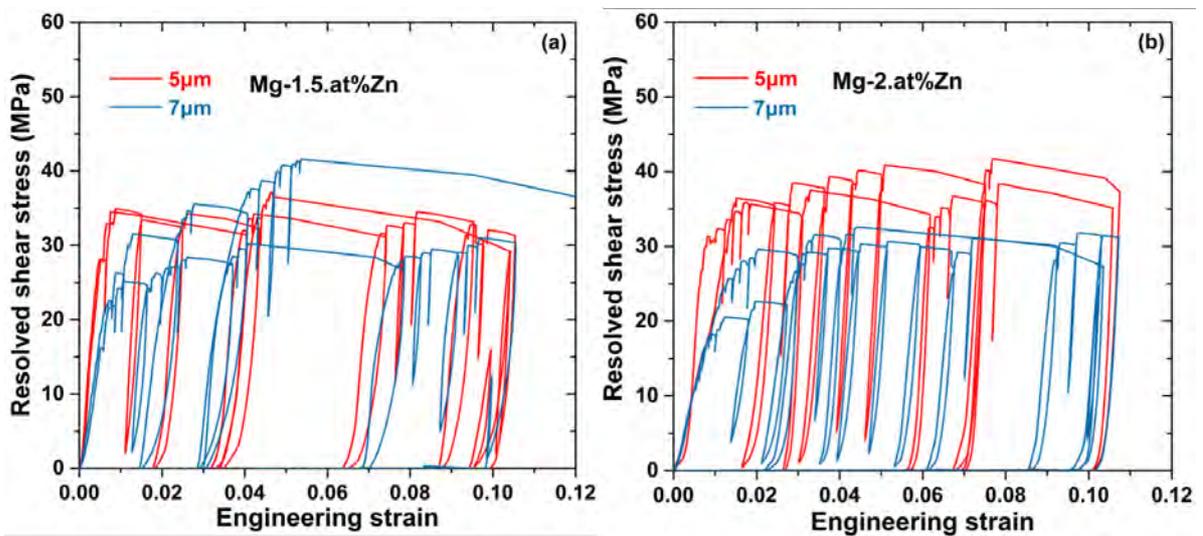

Fig. 5. Resolved shear stress ($\tau_{RSS}$) vs. engineering strain (ε) curves for micropillars of different dimensions tested at ambient temperature. (a) Mg-1.5 at.%Zn. and (b) Mg-2 at.%Zn.

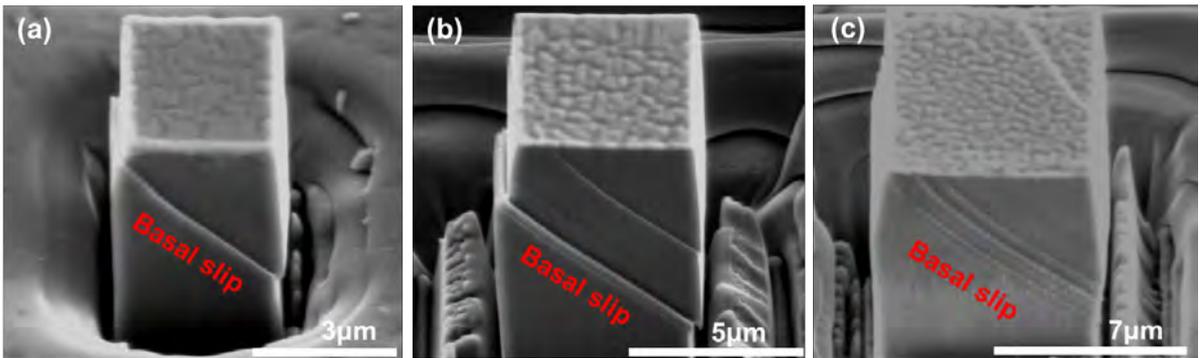

Fig. 6. SEM micrographs of the micropillars of pure Mg deformed up to ≈ 10%. (a) micropillar lateral dimension 3μm. (b) *Idem* 5μm. (c) *Idem* 7μm.

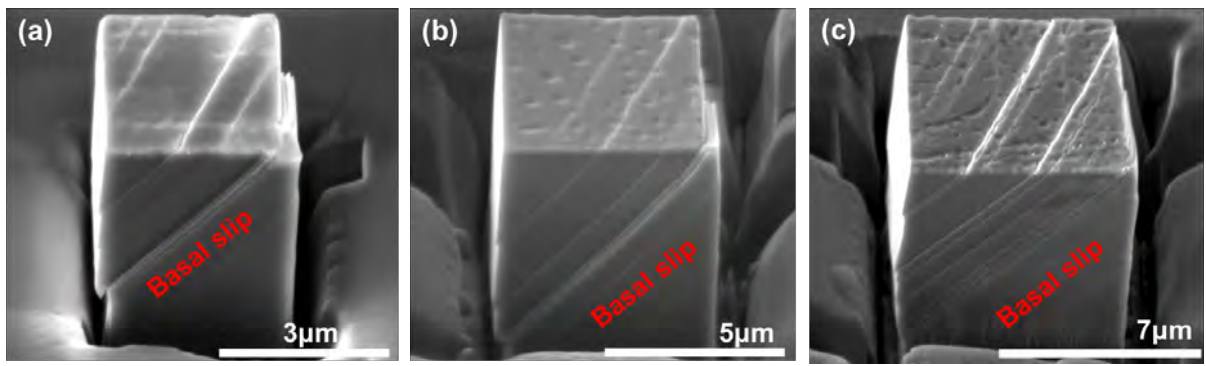

Fig. 7. SEM micrographs of the micropillars of Mg-9 at.%Al deformed up to ≈ 10%. (a) Micropillar lateral dimension 3μm. (b) *Idem* 5μm. (c) *Idem* 7μm.

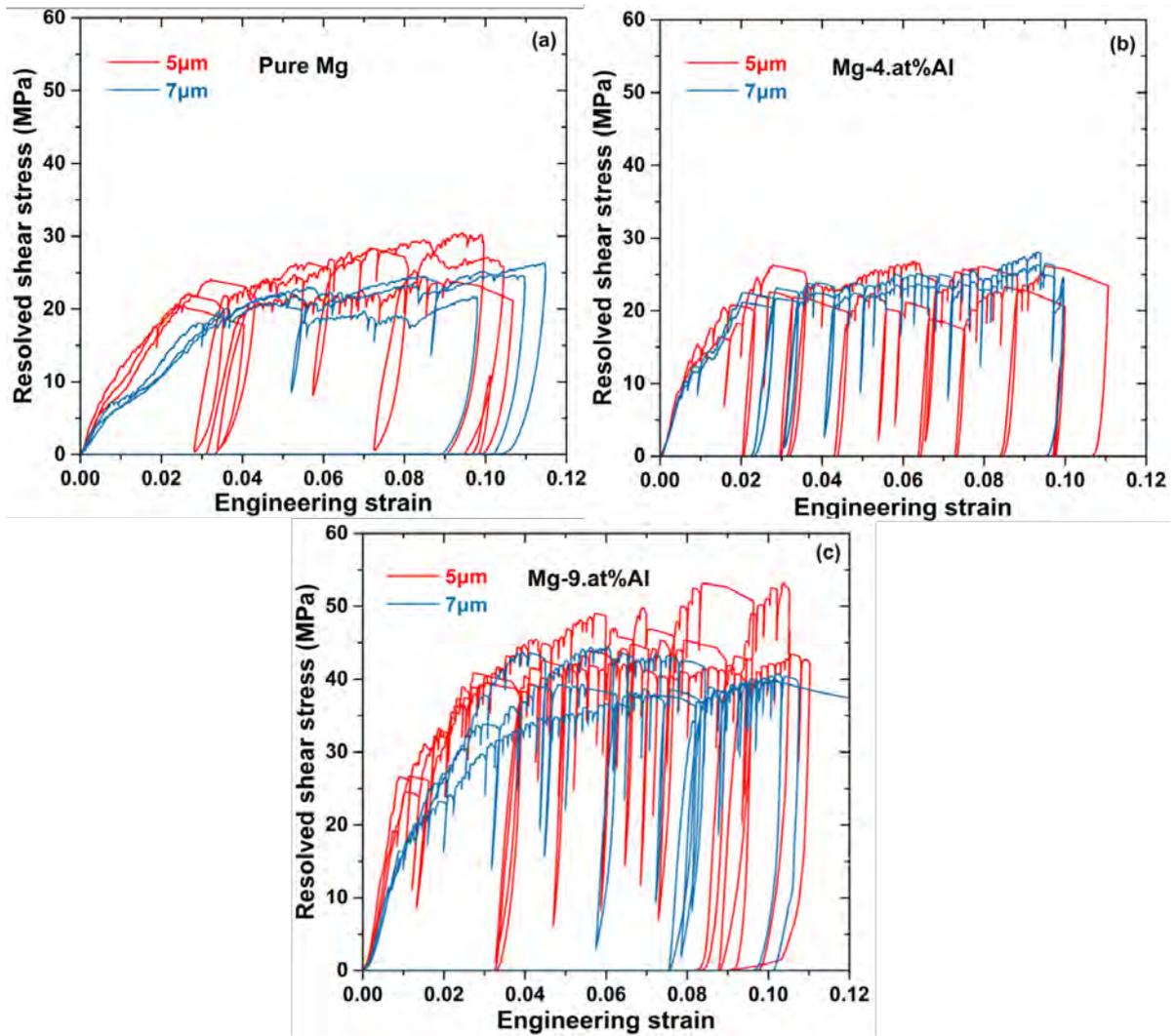

Fig. 8. Resolved shear stress and engineering strain curves obtained at 373 K (100 ºC) for (a) pure Mg; (b) Mg-4 at.%Al; and (c) Mg-9 at.%Al. The high temperature compression tests were performed on micropillars with lateral dimensions of 5 μm and 7 μm.

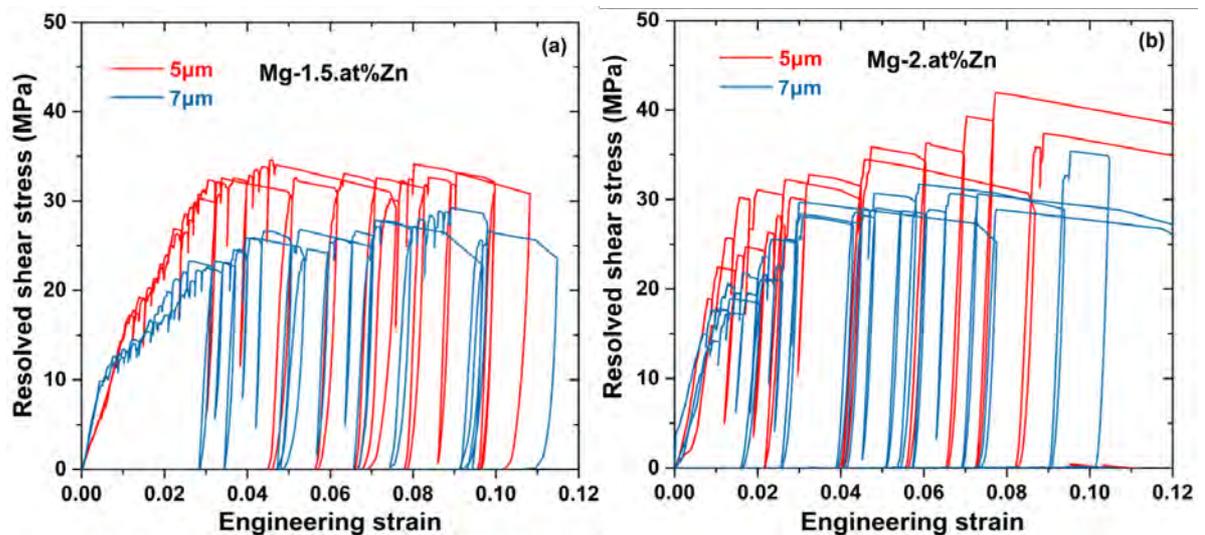

Fig. 9. Resolved shear stress and engineering strain curves obtained at 373 K (100 ºC) for (a) Mg-1.5 at.%Zn; and (b) Mg-2 at.%Zn.

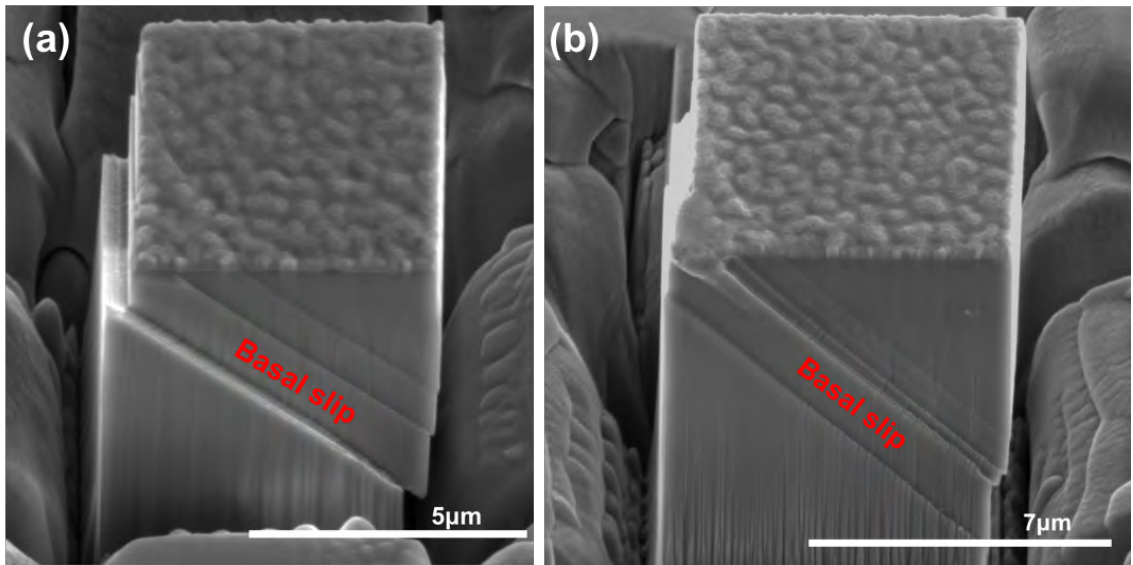

Fig. 10. SEM micrographs of the micropillars of the pure Mg deformed at 373 K up to ≈ 10%. (a) Micropillar lateral dimension 5μm; (b) *Idem* 7μm

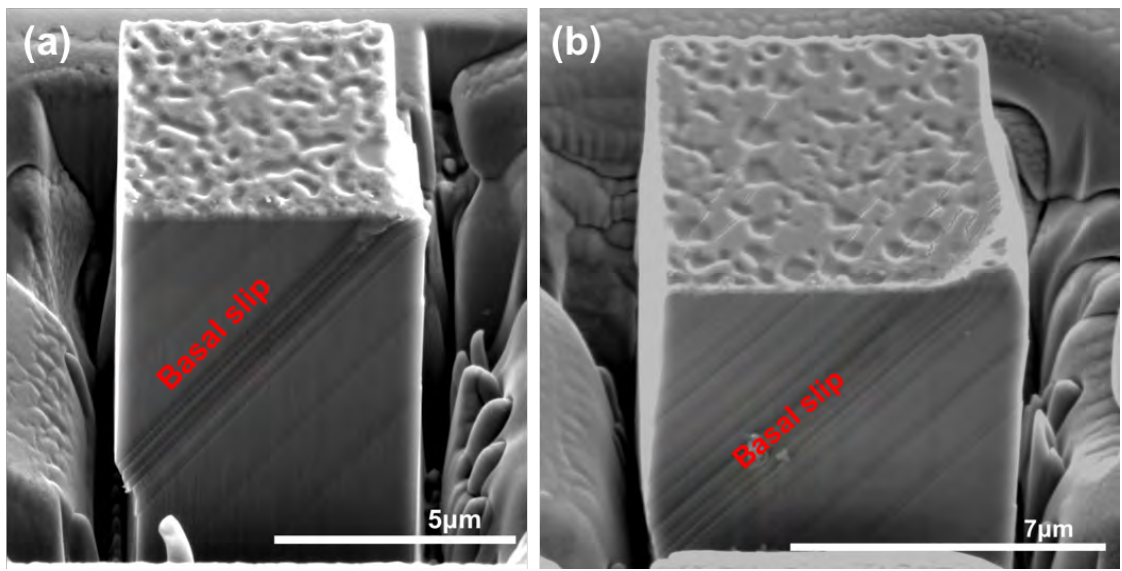

Fig. 11. SEM micrographs of the micropillars of Mg-9 at.%Al alloy deformed at 373 K up to ≈ 10%. (a) Micropillar lateral dimension 5μm; (b) *Idem* 7μm.

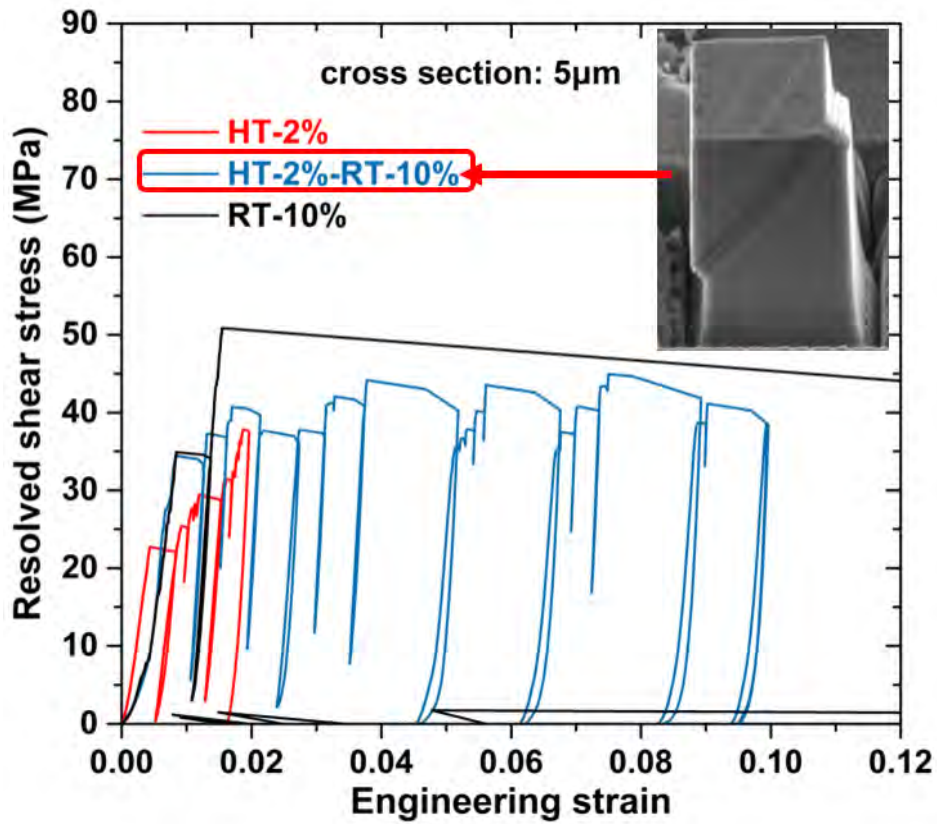

Fig. 12. The resolved shear stress ($\tau_{RSS}$)-strain ($\varepsilon$) curves of micropillars of 5 x 5 µm² cross section of the Mg-9 at.%Al alloy. The red curve represents the deformation up to 2% at 373 K; the blue curve corresponds to the deformation of the same micropillar up to 10% at the ambient temperature after the 2% pre-deformation at 373 K. The black curve shows the mechanical response of the micropillar deformed up to 10% strain at the room temperature. The inset shows the localization of the plastic deformation in bands in the micropillar deformed at ambient temperature after pre-deformation at 373 K.

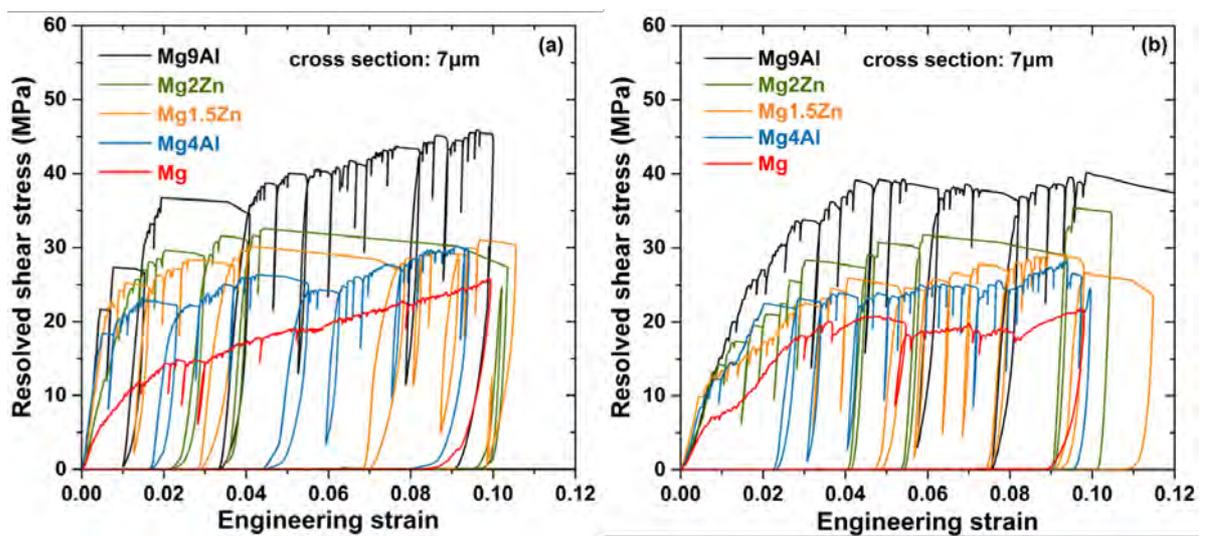

Fig. 13. Resolved shear stress ($\tau_{RSS}$) vs. engineering strain ($\varepsilon$) curves of Mg, Mg-4 at.%Al, Mg-1.5 at.%Zn, Mg-2 at.%Zn and Mg-9 at.%Al micropillars of 7 x 7 µm² cross section. (a) Ambient temperature. (b) 373 K (100 ºC).

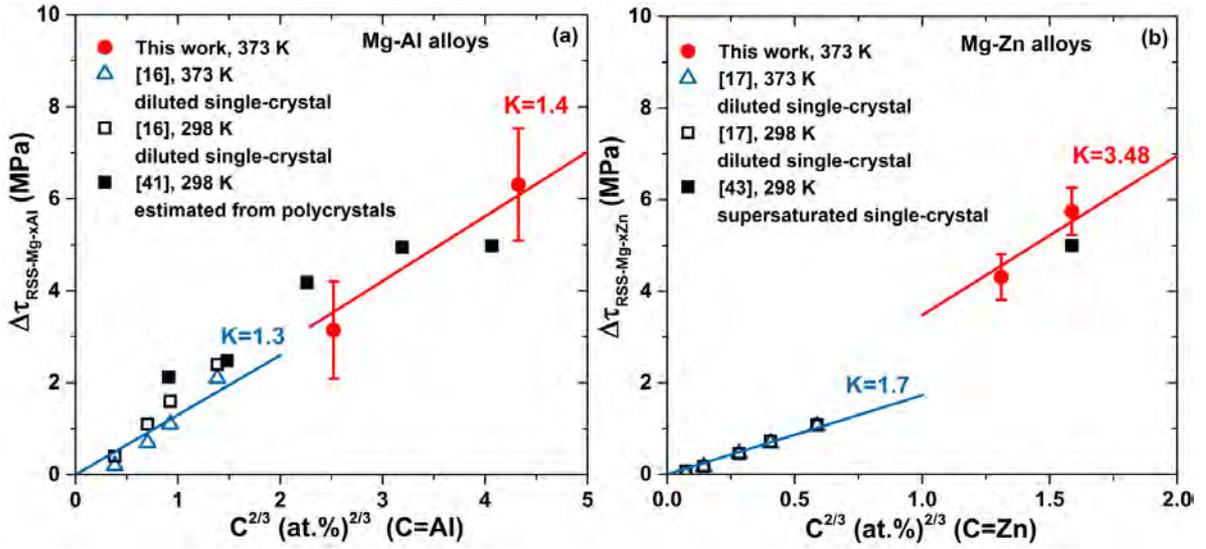

Fig. 14. The present experimental results of the solid strengthening on the increase of the CRSS, $\Delta\tau = \tau_{alloy} - \tau_{Mg}$, from compression tests in micropillars of 7 x 7 μm$^2$ cross-section at 373 K. (a) The effect of Al combined with the experimental data in the literature [16,41], and (b) the effect of Zn together with the experimental data [17,43].

Table 1. The Schmid factors for basal slip and tensile twinning for each selected grain

|  | Basal slip | Tensile twin |
|---|---|---|
| Pure Mg | 0.47 | 0.16 |
| Mg-4 at.%Al | 0.45 | 0.15 |
| Mg-9 at.%Al | 0.47 | 0.17 |
| Mg-1.5 at.%Zn | 0.48 | 0.19 |
| Mg-2 at.%Zn | 0.49 | 0.23 |

# Supplementary Material

In order to ascertain the dominant deformation mechanisms in more detail, a thin foil was extracted from the deformed 7x7 µm$^2$ micropillar tested at 373 K (100 ºC) in pure Mg and analyzed by the transmission EBSD. The SEM image of the deformed pillar is shown in Fig. 1a, which also includes the orientation of the thin lamella extracted from the deformed micropillar. The yellow dashed line in Fig. 1a indicates the position of the thin foil. The thin lamella included the whole longitudinal cross section of the micropillar and the inverse pole figure (IPF) obtained from the transmission EBSD analysis is presented in Fig. 1b and shows only one orientation for the micropillar section after deformation. Thus, twinning did not develop and basal slip was the only activated mechanism for the plastic deformation. Hence, it is reasonable to presume that the initial strain hardening was due to the interaction and exhaustion of basal dislocations as the pillar is compressed.

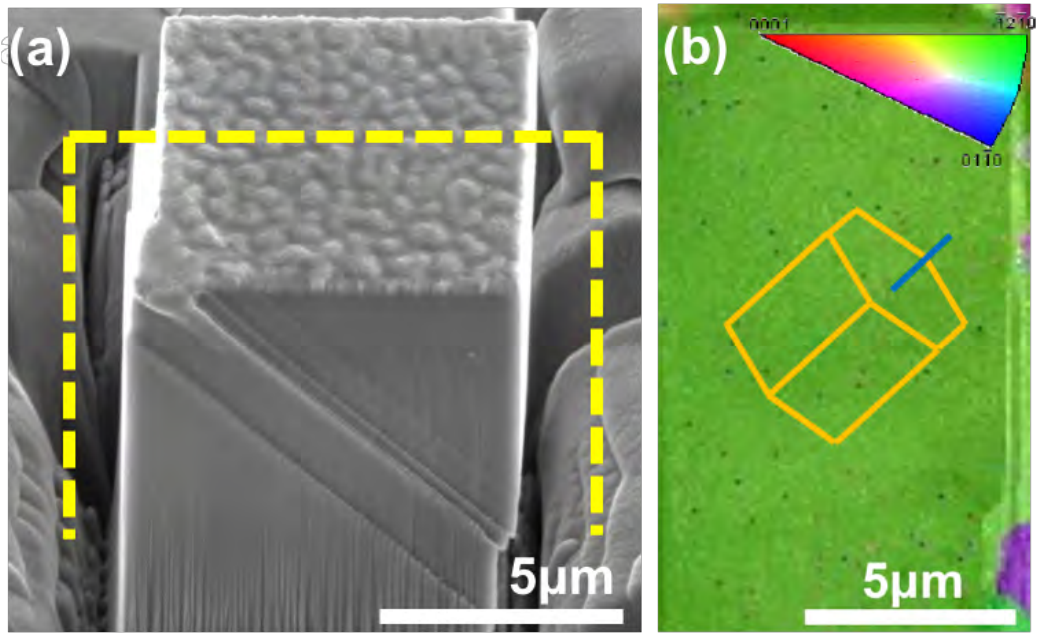

Fig. 1. (a) SEM micrograph of the 7 µm micropillar of pure Mg deformed at 373 K (100 ºC). The yellow dashed line shows the orientation of the thin lamella extracted from the micropillar. (b) IPF figure of the lamella obtained from the EBSD measurement.

In addition, the localization of the plastic deformation in slip bands can be observed in the movies of the in-situ micropillar compression tests at 273 K of the 3 µm (S1-Mg9Al-RT-3um.mp4) and 7 µm (S1-Mg9Al-RT-7um.mp4) micropillars in Mg-9 at.%Al alloy.